# A Novel Clustering Approach Based on Group Quasi-Consensus of Unstable Dynamic Linear High-Order Multi-Agent Systems

CAI, Ning[1] DIAO, Chen[1] KHAN, M. Junaid[2]

[1]College of Electrical Engineering, Northwest University for Nationalities, Lanzhou, China
[2]PN Engineering College, National University of Sciences and Technology, Islamabad, Pakistan

**Abstract:** This paper introduces a novel approach of clustering, which is based on group consensus of dynamic linear high-order multi-agent systems. The graph topology is associated with a selected multi-agent system, with each agent corresponding to one vertex. In order to reveal the cluster structure, the agents belonging to a similar cluster are expected to aggregate together. As theoretical foundation, a necessary and sufficient condition is given to check the group consensus. Two numerical instances are shown to illustrate the process of approach.

## 1. Introduction

As an unsupervised classification technique, clustering is important to be applied to various practical fields of data analysis, such as image segmentation [1-2], data mining [3], signal processing [4], community detection [5], gene expression study [6], and even linguistics simulation [7].

During the last several decades, different types of algorithms were developed to effectively solve the clustering problems, e.g. the graph partitioning algorithms [8], the G-N algorithms [5], the algorithms based on optimization of modularity functions [9], and the spectral clustering algorithms [10].

Besides the aforementioned types, there also exist certain clustering algorithms which are based on the analysis of network dynamics, including the random walk algorithms and the network synchronization algorithms. Random walk means moving along a random trail, each step to a next neighboring vertex. The random walk algorithms [11] endeavor to detect the clustering structure by imbuing a given network with flows of random walk. On the other hand, in a network synchronization algorithm [12], usually a specific Kuramoto oscillator model is concerned, and a synchronization of oscillation is expected to occur inside any cluster.

[1]Corresponding author: Cai, Ning (e-mail: caining91@tsinghua.org.cn)

In the current paper, a novel approach based on systems dynamics is proposed, in which a dynamic multi-agent system is associated with the graph topology concerned, with a one-to-one correspondence upon the vertices and the agents. The agents corresponding to the vertices that belong to a same cluster would be expected to automatically assemble together, thereby revealing the clustering configuration.

A relevant background of the current work is the group consensus problem that has been noticed by the control theory community since 2010. Yu and Wang earlier investigated group consensus phenomenon in [13]. Hu *et al.* studied the group consensus problem, both for systems with discontinuous inter-group information transmissions and hybrid protocols [14], and for systems with heterogeneous agents among different groups [15]. Su *et al.* considered both the adaptive pinning control scheme [16] and the case with multiple leaders [17] for the group synchronization problem of coupled oscillators. Xie *et al.* [18] gave some conditions for the group consensus of a class of first-order systems.

Our work is distinct from the other discussions about group consensus in the control theory literature. As far as our knowledge is concerned, all the existing articles address designing different information transmission protocols for the different anticipant groups, in order that the agents could form a presupposed clustering structure, which is already given beforehand. The agents are essentially heterogeneous to differentiate the clusters. However, in our scenario, there is no *a priori* knowledge about the clusters, and the clustering result is up to the topology of network. In this sense, the advantage of the current paper in practical utility is evident.

Except for proposing a novel clustering approach, the major contribution of the current paper also includes a necessary and sufficient condition for checking the group consensus of general LTI multi-agent systems, which extends the existing well-known condition for consensus.

Besides, the current paper provides an exemplification for the utilization of unstable dynamic systems, whereas in contrast, unstable systems are usually regarded as being insignificant by traditional control theory.

A lot of theoretical results on dynamic multi-agent systems have been highly developed in the area of control theory, especially about the consensus problem. However, the application instances corresponding to these theories that can well support them are seriously absent. Our exploration introduces a practical scenario from the field of data analysis, in order to apply, verify, enrich, or enlighten the relevant researches in control theory. This should be another contribution of the current paper.

The rest of this paper is organized as follows. In Section 2, an overview of the clustering approach is presented. The technical route of clustering based on dynamic LTI multi-agent systems is elaborated in Section 3, along with a necessary and sufficient condition for group consensus. Section 4 further illustrates the details of the approach via exhibiting two numerical instances. Finally, Section 4 includes the conclusion.

## 2. Problem Formulation

Suppose that the graph topology of the network concerned is given as $G = G(V, E)$, which is a graph with $N$ vertices, being undirected, weighted, and connected. Our purpose is to classify the vertex set $V$ into several distinct clusters, according to the topology of graph.

For this end, we propose a scheme based on dynamic multi-agent systems to achieve the clusters automatically. The scheme is summarized as follows.

First, select a multi-agent system to associate it with the graph topology as

$$\begin{cases} \dot{x}_1 = g(t, x_1) + \sum_{j=1}^{N} w_{1j} f(t, x_j, x_1) \\ \dot{x}_2 = g(t, x_2) + \sum_{j=1}^{N} w_{2j} f(t, x_j, x_2) \\ \quad \vdots \\ \dot{x}_N = g(t, x_N) + \sum_{j=1}^{N} w_{Nj} f(t, x_j, x_N) \end{cases} \tag{1}$$

where $x_i(t) \in X$ ($i \in \{1, 2, ..., N\}$) denotes the state of agent $i$ being varying with time, with $X$ a normed state space; $w_{ij} \in R^+$ denotes the edge weight of $G$ between vertices $i$ and $j$, with each vertex corresponding to a specific agent; and functions $g(\bullet)$ and $f(\bullet)$ denote the autonomous and interactive dynamics, respectively. In control theory, $f(\bullet)$ is also called the protocol for the communication of information.

Second, assign the initial values $x_i(0) \in X$ $(i = 1, 2, ..., N)$. If the structure and parameters of dynamic model (1) are appropriately set, then the states of agents would gradually achieve a group consensus or quasi-consensus, indicating the clustering result.

*Definition 1:* For system (1), if

$$\lim_{t\to\infty} \left\| x_i(t) - x_j(t) \right\| = 0$$

then agents *i* & *j* achieve an *agreement*. For a given vertex set $V_k \subset V$, if $\forall i, j \in V_k$, agents *i* & *j* achieve agreement, then the multi-agent system (1) achieves a *consensus* in $V_k$. If consensus is achieved in $V_1$, $V_2$, …, $V_\alpha$, respectively, with

$$V_1 \cup V_2 \cup \cdots \cup V_\alpha = V$$

then the overall system achieves a *group consensus*.

## 3. Dynamic Linear Multi-Agent Systems

Particularly, in this paper, an associated LTI multi-agent system model is adopted to solve the clustering problem, which is

$$\begin{cases} \dot{x}_1 = Ax_1 + F\sum_{j=1}^{N} w_{1j}(x_j - x_1) \\ \dot{x}_2 = Ax_2 + F\sum_{j=1}^{N} w_{1j}(x_j - x_1) \\ \quad \vdots \\ \dot{x}_N = Ax_N + F\sum_{j=1}^{N} w_{Nj}(x_j - x_N) \end{cases} \tag{2}$$

where $x_i(t) = \begin{bmatrix} x_{i1} & x_{i2} & \cdots & x_{id} \end{bmatrix}^T \in R^d$ $(i \in \{1, 2, ..., N\})$ denotes the state of agent *i*; and $A, F \in R^{d\times d}$ represent the autonomous and interactive dynamics, respectively.

*Remark 1:* The above model is general homogeneous high-order LTI multi-agent system, which can represent many real engineering systems such as multi-agent supporting system [19-20].

The key is to determine the parameter matrices *A* & *F*. Intuitively speaking, in order to obtain an acceptable clustering result, there should be three principles.

1. The multi-agent system should be homogeneous.
2. It is better that the system matrix *A* be unstable.
3. Some of the elements in $\{A - \lambda_2 F, A - \lambda_3 F, \cdots, A - \lambda_N F\}$ should be unstable and some be Hurwitz, where $\lambda_2, ..., \lambda_N$ are the nonzero eigenvalues of the Laplacian matrix *L*.

In the sequel, the three principles will be expounded in detail.

First, there is no *a priori* knowledge about the clustering. Thus, the agents along with the communication protocols must be identical and the system homogenous.

Second, let us review some of the earlier results about consensus.

*Lemma 1* [19-21]: For dynamic multi-agent system (2), if it achieves consensus, then as $t \to \infty$, the trajectory of each agent tends to be regulated by the equation $\dot{\xi} = A\xi$.

*Lemma 2* [22]: The Laplacian matrix $L$ of a directed graph $G$ has exactly a single zero eigenvalue $\lambda_1 = 0$ if and only if $G$ has a spanning tree, with the corresponding eigenvector $\phi = \begin{bmatrix} 1 & 1 & \cdots & 1 \end{bmatrix}^T$. Meanwhile, all the other eigenvalues $\lambda_2, ..., \lambda_N$ have positive real parts.

*Corollary 1:* The Laplacian matrix $L$ of an undirected graph $G$ has exactly a single zero eigenvalue $\lambda_1 = 0$ if and only if $G$ is connected, with the corresponding eigenvector $\phi = \begin{bmatrix} 1 & 1 & \cdots & 1 \end{bmatrix}^T$. Meanwhile, all the other eigenvalues $\lambda_2, ..., \lambda_N \in R^+$.

*Lemma 3* [19-21]: For a dynamic multi-agent system (2) with $\lambda_1 = 0, \lambda_2, \cdots, \lambda_N$ as the eigenvalues of the Laplacian matrix of the directed $G$, if $A$ is not Hurwitz, a necessary and sufficient condition for consensus is:

1) The graph topology $G$ has a spanning tree;
2) All the matrices $A - \lambda_i F$ ($i \in \{1, 2, ..., N\}$ $\lambda_i \neq 0$) are Hurwitz.

According to Lemma 1, the overall motion of any separate subsystem that achieves consensus is determined by *A*. If *A* is Hurwitz, then the consentaneous trajectories of different subsystems may coincide to converge to the origin, even no any communications exist. Therefore, *A* should not be Hurwitz. Since the ultimate purpose is to clearly differentiate between the clusters, matrix *A* is better to be unstable so that different subsystems could depart away from each other.

The following theorem should be important as a theoretical foundation for the group consensus of linear systems.

*Theorem 1:* For the multi-agent system (2), suppose that the spectrum of Laplacian matrix of the directed graph with spanning tree is

$$\{\lambda_1 = 0, \lambda_2, ..., \lambda_N\}$$

besides, the following matrices

$$A, A-\lambda_2 F, ..., A-\lambda_{h-1} F$$

are not Hurwitz, whereas

$$A-\lambda_h F, A-\lambda_{h+1} F, ..., A-\lambda_N F$$

are Hurwitz. A necessary and sufficient condition for the pair of agents *i* & *i*+1 (or *N* & 1 if *i*=*N*) to achieve agreement is that the *i*th row of matrix product (*PT*) takes the following structure:

$$\begin{bmatrix} \underset{(1)}{*} & \underset{(2)}{0} & \cdots & \underset{(h-1)}{0} & \underset{(h)}{*} & \cdots & \underset{(N)}{*} \end{bmatrix}$$

where $P \in R^{N\times N}$ denotes a feasible matrix that satisfies the equation

$$PL = \Gamma = \begin{bmatrix} 1 & -1 & & & \\ & 1 & -1 & & \\ & & \cdots & \cdots & \\ & & & 1 & -1 \\ -1 & 0 & \cdots & \cdots & 1 \end{bmatrix} \tag{3}$$

$T \in R^{N\times N}$ denotes a nonsingular matrix that transforms the Laplacian matrix into the similar Jordan canonical form:

$$T^{-1}LT = J = \begin{bmatrix} 0 & & & & \\ & \lambda_2 & * & & \\ & & \lambda_3 & \ddots & \\ & & & \ddots & * \\ & & & & \lambda_N \end{bmatrix}$$

and ‘$*$’ denotes any arbitrary value.

*Proof:* If the stack vector of the states of agents is defined as

$$x^T = \begin{bmatrix} x_1^T & x_2^T & \cdots & x_N^T \end{bmatrix}^T$$

then the system dynamics can be described by

$$\dot{x} = (I_N \otimes A - L \otimes F)x \tag{4}$$

Let $\tilde{x} = (T^{-1} \otimes I_d)x$, then (3) can be transformed into

$$\dot{\tilde{x}} = (I_N \otimes A - J \otimes F)\tilde{x} \tag{5}$$

which is equivalent to

$$\begin{bmatrix} \dot{\tilde{x}}_1 \\ \dot{\tilde{x}}_2 \\ \vdots \\ \vdots \\ \dot{\tilde{x}}_h \\ \vdots \\ \dot{\tilde{x}}_N \end{bmatrix} = \begin{bmatrix} A & & & & & & \\ & A-\lambda_2 F & \Omega & & & & \\ & & \ddots & \ddots & & & \\ & & & A-\lambda_{h-1}F & \Omega & & \\ & & & & A-\lambda_h F & \ddots & \\ & & & & & \ddots & \Omega \\ & & & & & & A-\lambda_N F \end{bmatrix} \begin{bmatrix} \tilde{x}_1 \\ \tilde{x}_2 \\ \vdots \\ \vdots \\ \tilde{x}_h \\ \vdots \\ \tilde{x}_N \end{bmatrix} \tag{6}$$

where '$\Omega$' denotes an indefinite matrix that may either be $F$ or zero. Define the auxiliary vectors

$$\eta_i = \sum\nolimits_{j=1}^{N} w_{ij}(x_j - x_i) \quad (i = 1, 2, \ldots, N)$$

and the stack of them as

$$\eta = \begin{bmatrix} \eta_1^T & \eta_2^T & \cdots & \eta_N^T \end{bmatrix}^T$$

It is evident that $\eta = (L \otimes I_d)x$, and

$$(P \otimes I_d)\eta = (PL \otimes I_d)x$$

Substituting (3) into the above equation yields

$$(PL \otimes I_d)x = (\Gamma \otimes I_d)x$$

It follows that

$$\begin{aligned} (\Gamma \otimes I_d)x &= (PL \otimes I_d)x \\ &= (PL \otimes I_d)(T \otimes I_d)\tilde{x} \\ &= (PTJT^{-1} \otimes I_d)(T \otimes I_d)\tilde{x} \\ &= (PTJ \otimes I_d)\tilde{x} \\ &= (PT \otimes I_d)(J \otimes I_d)\tilde{x} \end{aligned} \tag{7}$$

According to (6), $\dot{\tilde{x}}_N = (A - \lambda_N F)\tilde{x}_N$, thus $\lim_{t\to\infty} \tilde{x}_N(t) = 0$ because $A - \lambda_N F$ is Hurwitz. Also, $\dot{\tilde{x}}_{N-1} = (A - \lambda_{N-1}F)\tilde{x}_{N-1} + \Omega\tilde{x}_N$ stands, leading to $\lim_{t\to\infty} \tilde{x}_{N-1}(t) = 0$ because $A - \lambda_{N-1}F$ is Hurwitz and $\lim_{t\to\infty} \tilde{x}_N(t) = 0$. Such a similar analysis can be recursively conducted till $\tilde{x}_h$. As a result, it can be concluded that

$$\begin{cases} \lim_{t\to\infty} \tilde{x}_1(t) = * \\ \lim_{t\to\infty} \tilde{x}_2(t) = * \\ \quad\vdots \\ \lim_{t\to\infty} \tilde{x}_{h-1}(t) = * \\ \lim_{t\to\infty} \tilde{x}_h(t) = 0 \\ \quad\vdots \\ \lim_{t\to\infty} \tilde{x}_N(t) = 0 \end{cases}$$

Due to the structure of $J$ and the fact that $\lambda_1 = 0$, the limits of the first $d$

elements of vector $(J \otimes I_d)\tilde{x}$ are zero; the limits of the elements indexed from $d+1$ to $d(h-1)$ are indefinite; whilst the limits of the remaining elements are all zero.

According to the structure of $\Gamma$, the agents $i$ & $i$+1 ($i = 1, 2, ..., N-1$) achieving agreement implies that the elements of (7) indexed from $(i-1)d+1$ to $id$ tend to zero. These elements equal the products of the relevant row vectors of $(PT \otimes I_d)$ and the vector $(J \otimes I_d)\tilde{x}$. By noticing the fact that agreements are independent of the initial states and concerning the structure of $(J \otimes I_d)\tilde{x}$, one can easily conclude the requirement for $PT$ as the current theorem states.□

*Remark 2:* Although in the application scenario of the current paper, the graph of a system should be undirected, Theorem 1 still concerns the situation with a possibly directed graph and is thus more general.

*Remark 3:* A feasible *P* can be computed by solving the linear equation (3) through

$$P = \Gamma L^{\dagger}$$

where $L^{\dagger}$ denotes a Moore-Penrose inverse [23] of the singular Laplacian matrix *L*. Meanwhile, *T* is constituted by the generalized eigenvectors of *L*.

*Remark 4:* Both the matrices *P* and *T* are determined exclusively by the topology of graph. Therefore, the clustering result is determined by the graph topology, so long as a value of constant *h* is given.

*Remark 5:* Theorem 1 can be regarded as a bridge which connects consensus with the various non-consensus cases, till down to a completely unstable divergence. The smaller the value of *h*, the higher the degree of freedom that *PT* would possess to ensure agreement. If $h = 2$, then the overall system achieves a consensus, whatever the value of *PT* is. Oppositely, if all the matrices $A - \lambda_i F$ ($i \in \{1, 2, ..., N\}$) are not Hurwitz, then there should be no agreement. Actually, the existing well-known criteria for checking consensus [24-26] being analogous to Lemma 3 are just particular cases or corollaries of Theorem 1 as $h = 2$.

It is rational for Theorem 1 to consider only the case with a spanning tree, since the result is trivial about the opposite situation.

*Corollary 2:* For the multi-agent system (2), if the graph topology does not possess a spanning tree and *A* is not Hurwitz, then there will be no agreement between any two agents being respectively associated with different components of the graph.

# 4. Numerical Analysis

In this section, two simple numerical examples will be exhibited as follow to illustrate the process of clustering and validate the relevant theory.

*Example 1:* Consider a graph illustrated in Fig. 1.

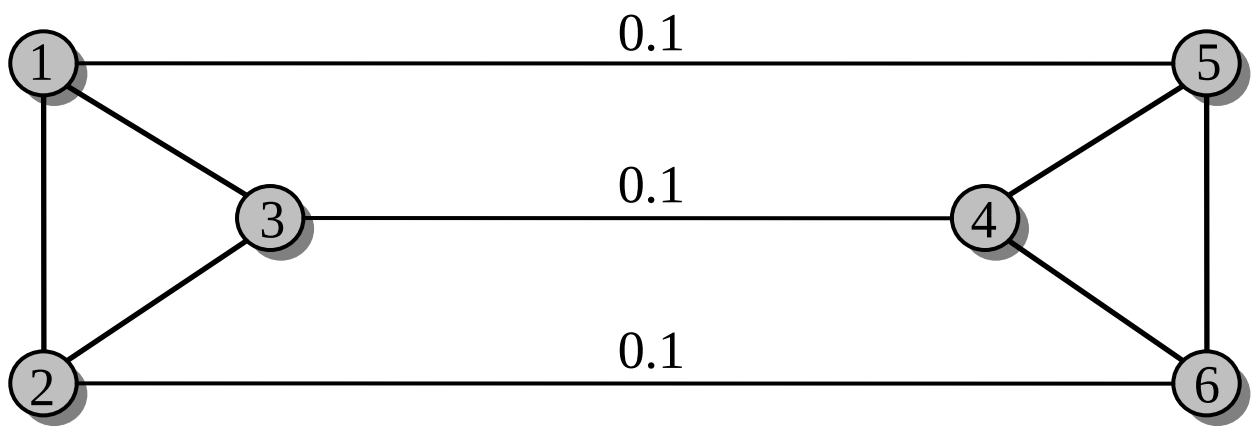

Fig. 1. First graph. Default edge weight is 1.

The adjacency matrix is

$$W = \begin{bmatrix} 0 & 1 & 1 & 0 & 0.1 & 0 \\ 1 & 0 & 1 & 0 & 0 & 0.1 \\ 1 & 1 & 0 & 0.1 & 0 & 0 \\ 0 & 0 & 0.1 & 0 & 1 & 1 \\ 0.1 & 0 & 0 & 1 & 0 & 0 \\ 0 & 0.1 & 0 & 1 & 0 & 0 \end{bmatrix}$$

and the Laplacian matrix is

$$L = \begin{bmatrix} 2 & -1 & -1 & 0 & -0.1 & 0 \\ -1 & 2.1 & -1 & 0 & 0 & -0.1 \\ -1 & -1 & 2.1 & -0.1 & 0 & 0 \\ 0 & 0 & -0.1 & 2.1 & -1 & -1 \\ -0.1 & 0 & 0 & -1 & 1.1 & 0 \\ 0 & -0.1 & 0 & -1 & 0 & 1.1 \end{bmatrix}$$

with the spectrum {0, 0.2, 1.095, 3, 3.105, 3.2}. Let the associated agents be LTI second-order systems with

$$A = \begin{bmatrix} 0.25 & 1 \\ -1 & 0.25 \end{bmatrix} \text{ and } F = \begin{bmatrix} 0 & -1 \\ 0.5 & 1 \end{bmatrix}$$

In this case, $A-\lambda_1 F$ and $A-\lambda_2 F$ are unstable, whereas $A-\lambda_3 F$ ,…, $A-\lambda_6 F$ are Hurwitz. Thus, $h=3$. It can be easily derived that

$$P = \begin{bmatrix} 0.3235 & -0.3235 & 0 & 0 & 0.0294 & -0.0294 \\ -0.0003 & 0.3232 & -0.3229 & -0.0104 & -0.0095 & 0.0199 \\ 1.5625 & 1.5625 & 1.8750 & -1.8750 & -1.5625 & -1.5625 \\ -0.0199 & 0.0095 & 0.0104 & 0.3229 & -0.6173 & 0.2944 \\ 0.0294 & -0.0294 & 0 & 0 & 0.9118 & -0.9118 \\ -1.8952 & -1.5423 & -1.5625 & 1.5625 & 1.2482 & 2.1893 \end{bmatrix}$$

and

$$T = \begin{bmatrix} 0.4082 & -0.4082 & 0.0352 & -0.2887 & -0.7062 & 0.2887 \\ 0.4082 & -0.4082 & -0.0352 & -0.2887 & 0.7062 & 0.2887 \\ 0.4082 & -0.4082 & 0 & 0.5774 & 0 & -0.5774 \\ 0.4082 & 0.4082 & 0 & 0.5774 & 0 & 0.5774 \\ 0.4082 & 0.4082 & 0.7062 & -0.2887 & 0.0352 & -0.2887 \\ 0.4082 & 0.4082 & -0.7062 & -0.2887 & -0.0352 & -0.2887 \end{bmatrix}$$

As a result

$$PT = \begin{bmatrix} 0 & 0 & 0.0643 & 0 & -0.4549 & 0 \\ 0 & 0 & -0.0322 & -0.2887 & 0.2274 & 0.2706 \\ 0 & -4.0825 & 0 & 0 & 0 & -0.3608 \\ 0 & 0 & -0.6450 & 0.2887 & -0.0113 & 0.2706 \\ 0 & 0 & 1.2899 & 0 & 0.0227 & 0 \\ 0 & 4.0825 & -0.6771 & 0 & 0.2161 & -0.1804 \end{bmatrix}$$

According to Theorem 1, it is now evident that agents 1~3 achieve a consensus, whereas agents 4~6 achieve another consensus. Therefore, a clustering result is successfully obtained. The state trajectories of one experiment are shown in Fig. 2.□

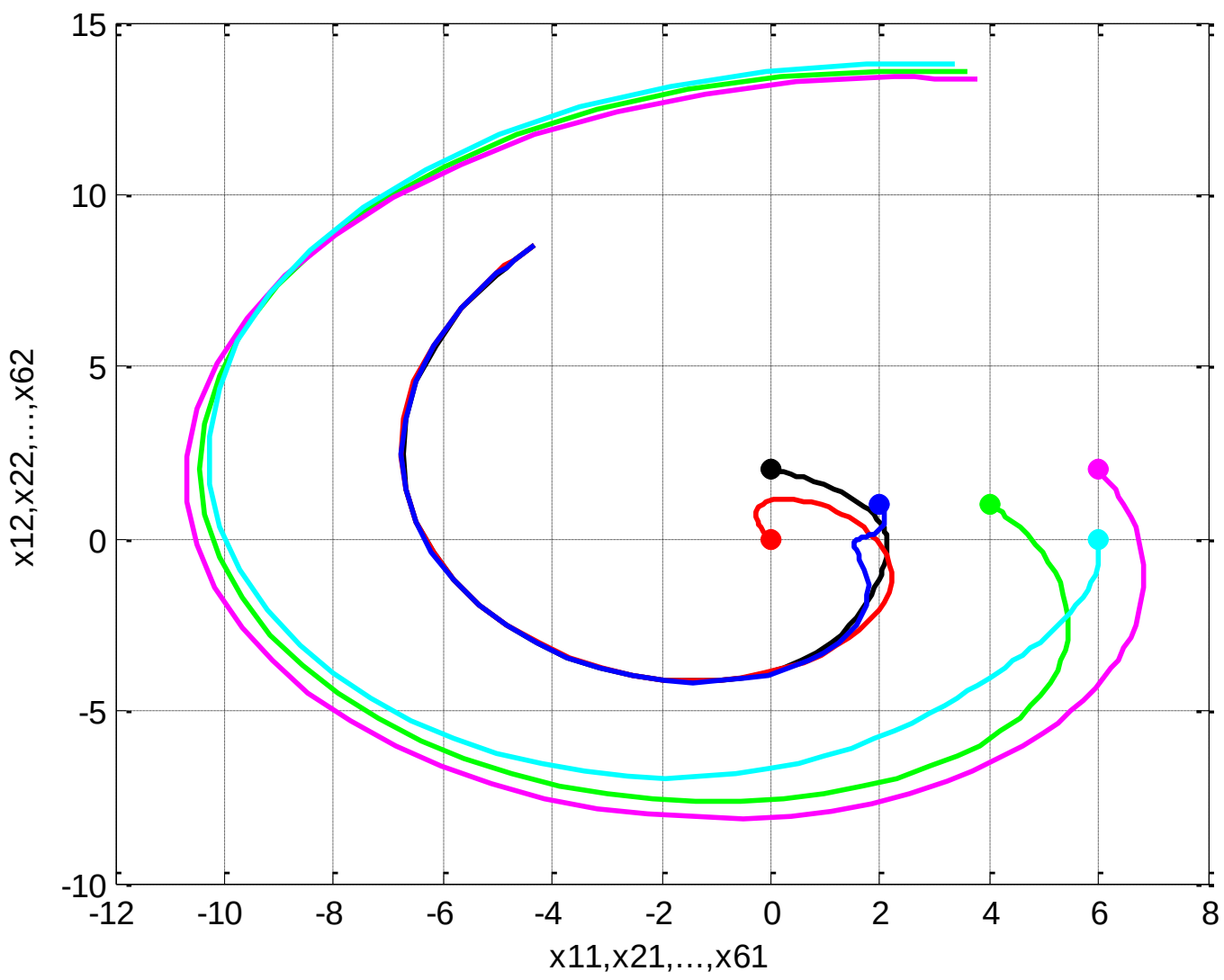

Fig. 2. Consentaneous state trajectories of Example 1 with $t \in [0,5]$.
Thick dots denote starting positions.

*Example 2:* In the second example, consider the graph illustrated in Fig. 3.

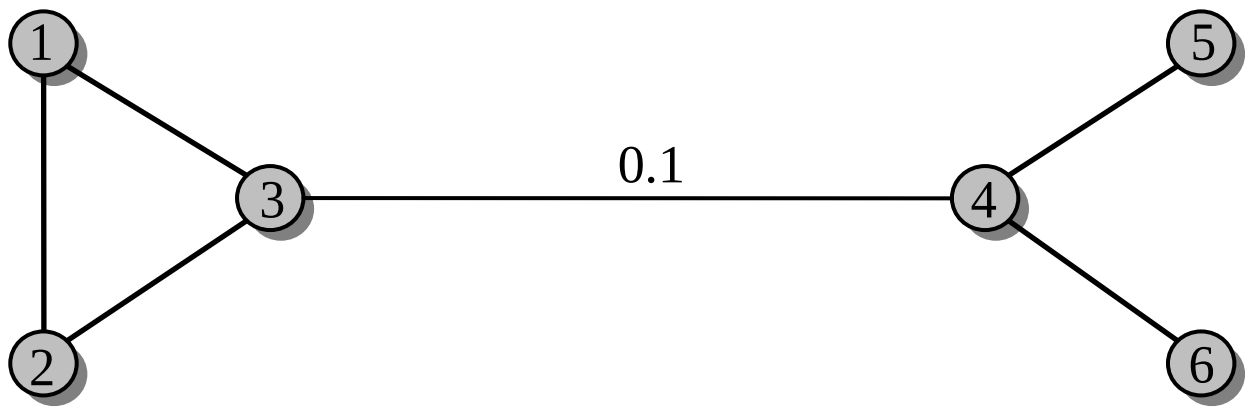

Fig. 3. Second graph. Default edge weight is 1.

The adjacency matrix is

$$W = \begin{bmatrix} 0 & 1 & 1 & 0 & 0 & 0 \\ 1 & 0 & 1 & 0 & 0 & 0 \\ 1 & 1 & 0 & 0.1 & 0 & 0 \\ 0 & 0 & 0.1 & 0 & 1 & 1 \\ 0 & 0 & 0 & 1 & 0 & 0 \\ 0 & 0 & 0 & 1 & 0 & 0 \end{bmatrix}$$

and the Laplacian matrix is

$$L = \begin{bmatrix} 2 & -1 & -1 & 0 & 0 & 0 \\ -1 & 2 & -1 & 0 & 0 & 0 \\ -1 & -1 & 2.1 & -0.1 & 0 & 0 \\ 0 & 0 & -0.1 & 2.1 & -1 & -1 \\ 0 & 0 & 0 & -1 & 1 & 0 \\ 0 & 0 & 0 & -1 & 0 & 1 \end{bmatrix}$$

with the spectrum {0, 0.0638, 1, 3, 3, 3.1362}. Let the associated agents be LTI second-order systems with

$$A = \begin{bmatrix} 0.25 & 2 \\ -2 & 0.25 \end{bmatrix} \text{ and } F = \begin{bmatrix} 0.5 & -1 \\ 4 & 0.5 \end{bmatrix}$$

In this case, $A-\lambda_1 F$ and $A-\lambda_2 F$ are unstable, whereas $A-\lambda_3 F$ ,…, $A-\lambda_6 F$ are Hurwitz. Still, $h=3$. It can be easily derived that

$$P = \begin{bmatrix} 0.3333 & -0.3333 & 0 & 0 & 0 & 0 \\ 0.1667 & 0.5 & -0.1667 & -0.1667 & -0.1667 & -0.1667 \\ 5 & 5 & 5 & -5 & -5 & -5 \\ -0.1667 & -0.1667 & -0.1667 & -0.1667 & -0.8333 & -0.1667 \\ 0 & 0 & 0 & 0 & 1 & -1 \\ -5.6667 & -5.3333 & -5 & 5 & 5 & 6 \end{bmatrix}$$

and

$$T = \begin{bmatrix} 0.4082 & 0.4169 & 0 & 0.2887 & 0.7071 & -0.276 \\ 0.4082 & 0.4169 & 0 & 0.2887 & -0.7071 & -0.276 \\ 0.4082 & 0.3903 & 0 & -0.5774 & 0 & 0.5896 \\ 0.4082 & -0.3903 & 0 & -0.5774 & 0 & -0.5896 \\ 0.4082 & -0.4169 & -0.7071 & 0.2887 & 0 & 0.276 \\ 0.4082 & -0.4169 & 0.7071 & 0.2887 & 0 & 0.276 \end{bmatrix}$$

As a result

$$PT = \begin{bmatrix} 0 & 0 & 0 & 0 & 0.4714 & 0 \\ 0 & 0.4169 & 0 & 0.2887 & -0.2357 & 0.276 \\ 0 & 12.2417 & 0 & 0 & 0 & 0.376 \\ 0 & 0.4169 & 0.7071 & -0.2887 & 0 & -0.276 \\ 0 & 0 & -1.4142 & 0 & 0 & 0 \\ 0 & -13.0755 & 0.7071 & 0 & -0.2357 & 0.176 \end{bmatrix}$$

According to Theorem 1, only two pairs of agents can achieve agreement, respectively, which are agents 1 & 2 and agents 5 & 6. The state trajectories of one experiment are shown in Fig. 4.

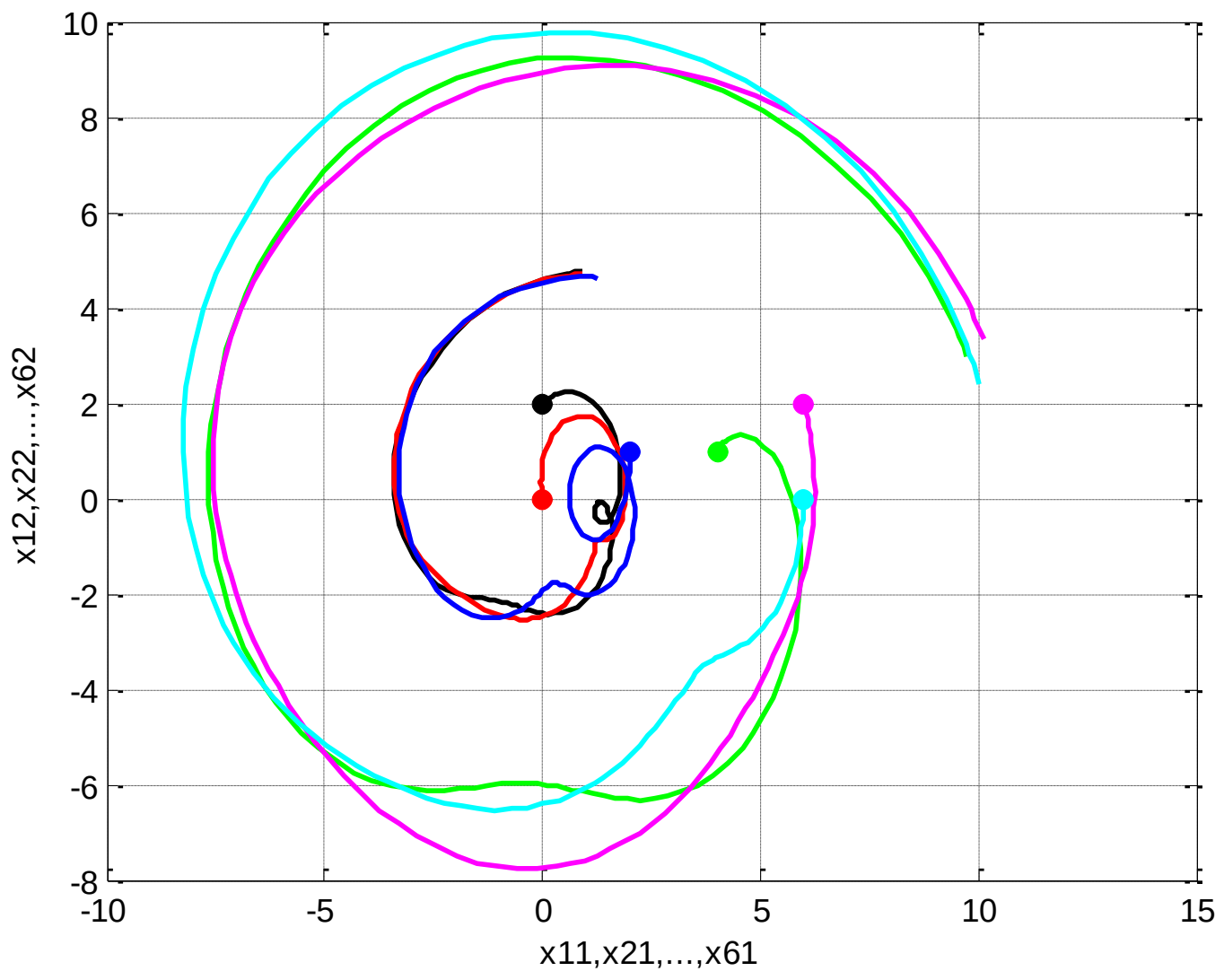

Fig. 4. Non-consentaneous state trajectories of Example 2 with $t \in [0,3]$.
Thick dots denote starting positions.

Although not all agents achieve precise group-consensus in this case, from Fig. 4, the six agents can still be practically differentiated into two distinct clusters, because the speed of divergence between the two clusters is much faster than the speed of divergence of the agents inside any cluster, during certain time span. The variation of the difference between agents 2 & 3 is illustrated in Fig. 5, which is $x_{22}(t) - x_{32}(t)$. □

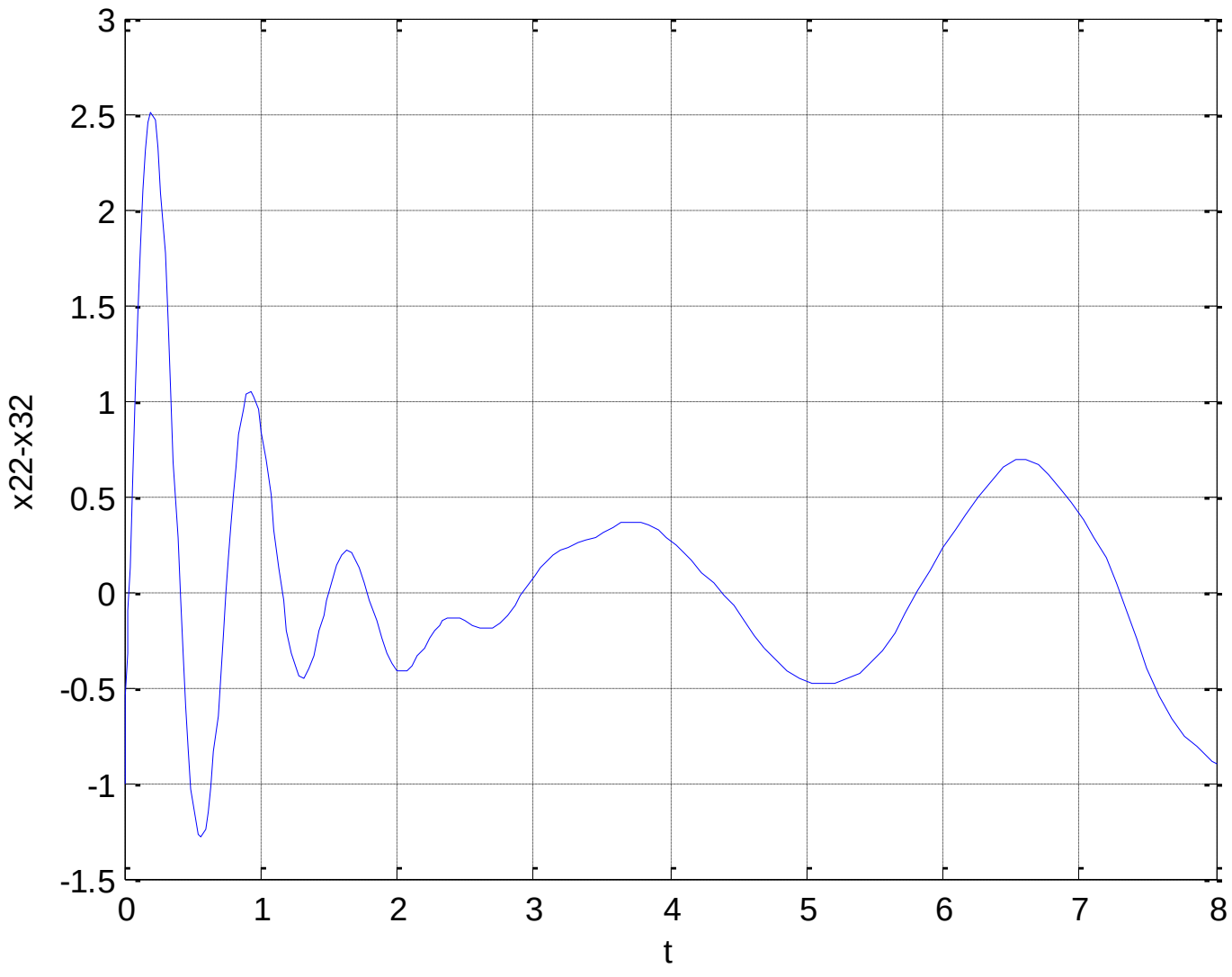

Fig. 5. Variation of difference between agents 2 & 3.

From Fig. 5, one can clearly sense that the vicissitude of the difference between agent 3 and the other agents in the same cluster is sophisticated, including two phases: before $t = 3$, there is a tendency of agreement; however afterwards, the tendency turns into a departure. This can be intuitively explained by a comparison with Example 1.

Let us survey the dynamic equation of agent 3, i.e.

$$\dot{x}_3 = Ax_3 + F\sum_{j\in V_1} w_{3j}(x_j - x_3) + F\sum_{j\in V_2} w_{3j}(x_j - x_3)$$

The vector field is composed of three components as shown above: the autonomous force, the interactive force from the nearest neighbors of the same cluster, and the interactive force from the other cluster. During the first phase of Example 2, the effect of third component is relatively weak, because of the small value of $w_{43}$, thus the states of agents belonging to a same cluster, i.e. 1, 2, & 3, tend to achieve a consensus; however, as time increases and the distance between the two clusters increases, the effect of the third component also keeps on increasing till ultimately overwhelms the diminished second component and tears agent 3 away from its nearest neighbors 1 & 2. On the other hand, as to Example 1, a precise group-consensus can be achieved because the effect from the third component is balanced upon every agent.

## 6. Conclusion

This paper proposes a novel approach of clustering, which is based on the group consensus of dynamic high-order LTI multi-agent systems. In order to classify the vertices of a given graph into clusters, the graph is associated with an appropriate multi-agent system. It is expected that the agents corresponding to the same cluster tend to automatically aggregate together in the state space, whereas different clusters depart away from each other. A necessary and sufficient condition of group consensus is presented for LTI homogeneous systems, expanding the earlier condition of consensus. The utility of this condition ought to be broader, beyond the clustering problem here. Two simple numerical instances are shown to illustrate the detail of the approach. It is evident that several potential directions exist for the future research: 1) the current approach can be applied to solve practical problems such as image segmentation; 2) the approach could be extended to the more complicated cases, e.g. with time-varying graph topology or nonlinear high-order communication protocol; 3) the mechanism of group consensus or quasi-consensus can be further investigated in depth, e.g. a more straight relationship between the geometric characteristic of graph topology and group consensus might possibly be uncovered.

## Acknowledgments

This work is supported by Program for Young Talents of State Ethnic Affairs Commission (SEAC) China (Grant [2013] 231), and by National Natural Science Foundation (NNSF) of China (Grants 61263002 & 61374054).

## References

[1] A. Rajendran, R. Dhanasekaran, "Enhanced possibilistic fuzzy C-Means algorithm for normal and pathological brain tissue segmentation on magnetic resonance brain image", *Arab. J. Sci. Eng.*, vol. 38, pp. 2375-2388, 2013.

[2] S. Zeng, R. Huang, Z. Kang, N. Sang, "Image segmentation using spectral clustering of Gaussian mixture models", *Neurocomputing*, vol. 144, pp. 346-356, 2014.

[3] U. Fayyad, G. Piatetsky-Shapiro, P. Smyth, "From data mining to knowledge discovery in databases", *AI magazine*, vol. 17, pp. 37-54, 1996.

[4] F. Mohamad, T. S. Cevat, A. Mehdi, P. Farzad, "Fracture characteristics of AISI D2 tool steel at different tempering temperatures using acoustic emission and fuzzy C-Means clustering", *Arab. J. Sci. Eng.*, vol. 38, pp. 2205-2217, 2013.

[5] M. Girvan, M. E. J. Newman, "Community structure in social and biological networks", *Proc.Natl. Acad. Sci. USA*, vol. 99, pp. 7821-7826, 2002.

[6] D. Jiang, C. Tang, A. Zhang, "Cluster analysis for gene expression data: A survey", *IEEE Trans. Knowl. Data Eng.*, vol. 16, pp. 1370-1386, 2004.

[7] A. Cangelosi, D. Parisi (Eds), *Simulating the Evolution of Language*, London: Springer-Verlag, 2002.

[8] D. G. Corneil, C. C. Gotlieb, "An efficient algorithm for graph isomorphism", *J. ACM*, vol. 17, pp. 51-64, 1970.

[9] M. E. J. Newman, "Modularity and community structure in networks", *Proc.Natl. Acad. Sci. USA*, vol. 103, pp. 8577-8582, 2006.

[10] F. R. K. Chung, *Spectral Graph Theory*, American Mathematical Society, 1997.

[11] S. M. von Dongen, *Graph Clustering by Flow Simulation*, Doctoral Dissertation, University of Utrecht, 2000.

[12] A. Arenas, A. Diaz-Guilera, C. J. Perez-Vicente, "Synchronization reveals topological scales in complex networks", *Phys. Rev. Lett.*, vol. 96, 2006.

[13] J. Yu, L. Wang, "Group consensus of multi-agent systems with switching topologies and communication delays", *Syst. Control Lett.*, vol. 59, pp. 340-348, 2010.

[14] H. Hu, L. Yu, W. Zhang *et al.*, "Group consensus in multi-agent systems with hybrid protocol", *J. Franklin Inst.*, vol. 350, pp. 575-597, 2013.

[15] H. Hu, W. Yu, Q. Xuan *et al.*, "Group consensus for heterogeneous multi-agent systems with parametric uncertainties", *Neurocomputing*, vol. 142, pp. 383-392, 2014.

[16] H. Su, Z. Rong, M. Chen *et al.*, "Decentralized adaptive pinning control for cluster synchronization of complex dynamical networks", *IEEE Trans. Cybern.*, vol. 43, pp. 394-399, 2013.

[17] H. Su, M. Chen, X. Wang *et al.*, "Adaptive cluster synchronization of coupled harmonic oscillators with multiple leaders", *IET Control Theory Appl.*, vol. 7, pp. 765-772, 2013.

[18] D. Xie, Q. Liu, L. Lv *et al.*, "Necessary and sufficient condition for the group consensus of multi-agent systems", *Appl. Math. Comput.*, vol. 243, pp. 870-878, 2014.

[19] N. Cai, *Swarm Stability and Controllability of High-Order Swarm Systems*, Doctoral Dissertation, Tsinghua University, 2010. (In Chinese)

[20] N. Cai, J. Xi, Y. Zhong, "Asymptotic swarm stability of high order dynamical multi-agent systems: Condition and application", *Control Intell. Syst.*, vol. 40, pp. 33-39, 2012.

[21] N. Cai, J. Xi, Y. Zhong, "Swarm stability of high order linear time-invariant swarm systems", *IET Control Theory Appl.*, vol. 5, pp. 402-408, 2011.

[22] W. Ren, R. Beard, "Consensus seeking in multiagent systems under dynamically changing interaction topologies", *IEEE Trans. Autom. Control*, vol. 50, pp. 655-661, 2005.

[23] R. A. Horn, C. R. Johnson, *Matrix Analysis*, Cambridge: Cambridge University Press, 1985.

[24] F. Xiao, L. Wang, "Consensus problems for high-dimensional multi-agent systems", *IET Control Theory Appl.*, vol. 1, pp. 830-837, 2007.

[25] J. Wang, D. Cheng, X. Hu, "Consensus of multi-agent linear dynamic systems", *Asian J. Control*, vol. 10, pp. 144-155, 2008.

[26] Z. Li, Z. Duan, G. Chen *et al.*, "Consensus of multiagent systems and synchronization of complex networks: A unified viewpoint", *IEEE Trans Circuit. Syst. I: Regular Papers*, vol. 57, pp. 213-224, 2010.